\documentclass[a4paper,11pt]{article}

\pdfoutput=1
\usepackage{amsmath}
\usepackage{graphicx}

\usepackage[english]{babel}
\usepackage[utf8]{inputenc}
\usepackage{pdflscape}
\usepackage{enumerate}
\usepackage{amsbsy}
\usepackage{amsmath} %used by default
\usepackage{graphics}
\usepackage{mathrsfs}
\usepackage{wrapfig}
\usepackage{mathtools}

\usepackage{amsfonts}
\usepackage{pstricks}
\usepackage{color}
\usepackage{setspace}

\usepackage{jcappub_ver} % jcappub_ver load hyperref, load it last

\newcommand{\bea}{\begin{eqnarray}} \newcommand{\eea}{\end{eqnarray}}
\newcommand{\el}{\nonumber \\}
\newcommand{\re}[1]{(\ref{#1})}

\newcommand{\pat}{\partial}

\renewcommand{\sec}[1]{section \ref{#1}}

\newcommand{\para}{\paragraph}

\renewcommand{\a}{\alpha}
\renewcommand{\b}{\beta}
\renewcommand{\c}{\gamma}
\renewcommand{\d}{\delta}

\newcommand{\ha}{\frac{1}{2}}
\newcommand{\erf}{\mathrm{erf}}

\newcommand{\rmd}{\mathrm{d}}

\newcommand{\av}[1]{\langle{#1}\rangle}
\newcommand{\Mpl}{M_{{}_{\mathrm{Pl}}}}

\newcommand{\lc}{\lambda_\text{c}}
\newcommand{\thetat}{\tilde{\theta}}
\newcommand{\sigmat}{\tilde{\sigma}}

\title{The effect of matter discreteness on gravitational wave propagation in post-geometrical optics}

\author[a]{Sena Atli}
\author[a, b]{and Syksy R\"{a}s\"{a}nen}

\affiliation[a]{University of Helsinki, Department of Physics,\\ P.O. Box 64, FIN-00014 University of Helsinki, Finland}

\affiliation[b]{University of Helsinki, Helsinki Institute of Physics,\\ P.O. Box 64, FIN-00014 University of Helsinki, Finland}

\emailAdd{senaxatli@gmail.com}
\emailAdd{syksy.rasanen@iki.fi}

\abstract{
The gravitational wave equation of motion includes direct coupling to the Riemann tensor. The curvature terms are usually neglected, but they can be large at the location of matter particles and impact the angular diameter distance. We apply the recently introduced post-geometrical optics approximation that includes curvature to gravitational wave propagation. Assuming that particles are localised within their Compton wavelength, the curvature due to electrons leads to a large effect on the angular diameter distance, but caustic formation invalidates the post-geometrical optics approximation. We conclude that the interesting regime of validity of the approximation is limited, as it ceases to apply when the curvature effects become large. Other methods are needed to evaluate the effect of curvature spikes, and the localisation of particles due to decoherence also needs further work. 
}

\begin{document}

\begin{flushleft}
	\hfill		 HIP-2025-26/TH \\
\end{flushleft}
 
\setcounter{tocdepth}{3}

\setcounter{secnumdepth}{3}

\maketitle

\section{Introduction} \label{sec:intro}

\para{Large curvature from small particles.}

The propagation of gravitational waves is often discussed in the geometrical optics approximation, where they follow null geodesics \cite{Isaacson:1968hbi}, \cite{Maggiore:2007ulw} (page 42). It is well known that this approximation breaks down in lensing when the Schwarzschild radius of the lens is shorter than the wavelength of the gravitational wave, and wave optics is used instead \cite{Ohanian:1973, Ohanian:1974}. (See \cite{Braga:2024pik} for a treatment of the geometrical optics and wave optics approximations in the path integral formalism.) Neither geometrical optics nor wave optics includes the effect of direct coupling to spacetime curvature in the gravitational wave equation of motion \cite{Lanczos:1925, Peters:1966, Isaacson:1968hbi}. As the mean curvature of the universe and the curvature due to macroscopic objects like stars and galaxies are small compared to the energy of gravitational waves, these terms are usually neglected. However, when the microscopic structure of matter is taken into account, the curvature can be large. For matter consisting of individual well-separated particles, the curvature has high spikes at the locations of the particles and is negligible between them. If a particle of mass $m$ is localised within its Compton wavelength, its energy density is $\rho\sim10^{-3} m^4$ and the spacetime curvature is $\sim\rho/\Mpl^2$.

In \cite{Koksbang:2021alx} a post-geometrical approximation was introduced to treat the effect of the discreteness of dark matter particles on electromagnetic waves. The effect on redshift was found to be negligible. In contrast, the effect on the angular diameter distance was found to be possibly large, as it depends quadratically on the matter density and its derivatives. In addition to corrections proportional to $\rho/(\Mpl^2 E^2)$ and $\rho^2/(\Mpl^4 E^4)$, where $E$ is the energy of the wave, gradients of the density lead to a term proportional to $m^2 \rho^2/(\Mpl^2 E^4)$. It was noted that this can have a significant effect on the angular diameter distance to the cosmic microwave background (CMB), and as this has not been observed, dark matter masses $10^4$ GeV $\gtrsim m\gtrsim100$ MeV are excluded, assuming the validity of the treatment.

Gravitational waves are more sensitive to curvature than electromagnetic waves, as their wavelengths are longer and energies smaller. They are also affected by the curvature spikes of baryons and electrons in addition to dark matter, as gravitational waves pass through charged particles with negligible interaction, unlike electromagnetic radiation. The smallest frequency of  the gravitational waves detected by the LIGO--Virgo--KAGRA network is of the order 10 Hz, corresponding to wavelength $10^7$ m and energy $10^{-14}$ eV. Naively applying the result found for electromagnetic waves, the correction to the distance is larger than unity for $m\gtrsim10$ eV, and the approximation is valid for $m\lesssim100$ MeV. We do the calculation for gravitational waves to check these estimates and look for new features compared to the electromagnetic case.

In \sec{sec:post-geometric} we apply the post-geometrical approximation developed in \cite{Koksbang:2021alx} to gravitational waves. We find that if particles are localised within their Compton wavelength, the curvature due to electrons has a large impact, leading to formation of caustics that invalidate the approximation. In \sec{sec:conc} we summarise our findings and highlight open questions. In appendix \ref{app:turn} we show that when the induced graviton mass is small, the change in wavefront direction is also small.

\section{Post-geometrical optics} \label{sec:post-geometric}

\subsection{Post-geometrical approximation}

\para{The equation of motion.}

We mostly use the formalism and notation introduced in \cite{Koksbang:2021alx}. We consider the propagation of gravitational waves described by the metric perturbation $h_{\a\b}$ around an arbitrary background metric $g_{\a\b}$ to linear order in $h_{\a\b}$. We start from the Einstein equation
\bea \label{Einstein}
  R_{\a\b} - \ha R g_{\a\b} &=& \Mpl^{-2} T_{\a\b} \ ,
\eea
where $R_{\a\b}$ is the Ricci tensor, $R\equiv g^{\a\b} R_{\a\b}$, $\Mpl$ is the reduced Planck mass, and $T_{\a\b}$ is the energy-momentum tensor. We get the equation of motion of gravitational waves by shifting $g_{\a\b}\to g_{\a\b}+h_{\a\b}$ and expanding \re{Einstein} to first order in $h_{\a\b}$. Under this shift, the change in the Ricci tensor is $\d R_{\a\b}=\nabla_\c \d\Gamma^{\c}_{\a\b} - \nabla_\b \d\Gamma^{\c}_{\c\a}$, where $\d\Gamma^{\c}_{\a\b}=\nabla_{(\a} h_{\b)}{}^\c-\ha \nabla^\c h_{\a\b}$ is the change in the connection. Inserting this into \re{Einstein} and noting that $\d g^{\a\b}=-h^{\a\b}$ (all indices are lowered and raised with $g_{\a\b}$ and its inverse), we obtain \cite{Lanczos:1925, Peters:1966, Isaacson:1968hbi}
\bea \label{eom}
  - 2 \Mpl^{-2} \d T_{\a\b} &=& \Box \bar h_{\a\b} - 2 \nabla_{(\a} \nabla^\mu \bar h_{\b)\mu} + \nabla_\mu \nabla_\nu \bar h^{\mu\nu} g_{\a\b}  + 2 R_{\a\mu\b\nu} \bar h^{\mu\nu} - 2 R_{(\a}{}^\mu \bar h_{\b)\mu} \el
  && + R \bar h_{\a\b} - R_{\mu\nu} \bar h^{\mu\nu} g_{\a\b} \el
  &=& \Box \bar h_{\a\b} - 2 \nabla_{(\a} \nabla^\mu \bar h_{\b)\mu} + \nabla_\mu \nabla_\nu \bar h^{\mu\nu} g_{\a\b}  + 2 C_{\a\mu\b\nu} \bar h^{\mu\nu} - 4 R_{(\a}{}^\mu \bar h_{\b)\mu} + \frac{4}{3} R \bar h_{\a\b} \el
  && + \bar h \left( R_{\a\b} - \frac{1}{3} R g_{\a\b} \right) \el
  &=& \Box \bar h_{\a\b} + 2 C_{\a\mu\b\nu} \bar h^{\mu\nu} - 4 R_{(\a}{}^\mu \bar h_{\b)\mu} + \frac{4}{3} R \bar h_{\a\b} + \bar h \left( R_{\a\b} - \frac{1}{3} R g_{\a\b} \right) \ ,
\eea
where $\Box\equiv g^{\a\b}\nabla_\a \nabla_\b$, $\bar h_{\a\b}\equiv h_{\a\b}-\ha g_{\a\b} h$ (with $h\equiv g^{\a\b} h_{\a\b}$) is the trace-reversed metric perturbation, $R_{\a\mu\b\nu}$ is the Riemann tensor, and $C_{\a\mu\b\nu}$ is the Weyl tensor. In the third equality we have adopted the Lorenz gauge (also called the generalised De Donder gauge) defined by
\bea \label{Lorenz}
  \nabla^\mu \bar h_{\mu\a} = 0 \ .
\eea
The equation of motion \re{eom} applies to any metric perturbation. We are interested in gravitational waves, described by the traceless part of $\bar h_{\a\b}$. In vacuum, the residual gauge freedom of the Lorenz gauge can be used to set $\bar h=0$ without loss of generality, so the trace does not propagate outside matter, and describes parts of the metric other than gravitational waves. For discussion of the different gravitational and matter modes for gravitational waves passing through matter, see \cite{Gayer:1979ff, Ehlers:1987nm, Ehlers:1996}. Inside matter $\bar h$ is in general non-zero. We do not consider the generation of gravitational waves and neglect the impact of gravitational waves on matter, so we put $\d T^\a{}_{\b}=0$, which is equivalent to $\d T_{\a\b}=h_{\mu(\a} T^\mu{}_{\b)}$. Inserting this into \re{eom} and using the Einstein equation \re{Einstein}, we obtain
\bea \label{eom2}
  0 &=& \Box \bar h_{\a\b} + 2 C_{\a\mu\b\nu} \bar h^{\mu\nu} - 2 R_{(\a}{}^\mu \bar h_{\b)\mu} + \frac{1}{3} R \bar h_{\a\b} +  \frac{1}{6} \bar h R g_{\a\b} \equiv \Box \bar h_{\a\b} + M_{\a\b}{}^{\mu\nu} \bar h_{\mu\nu} \ ,
\eea
where we have defined
\bea \label{M}
  M_{\a\b}{}^{\mu\nu} &\equiv& 2 C_\a{}^{(\mu}{}_\b{}^{\nu)} - 2 R_{(\a}{}^{(\mu} \d_{\b)}{}^{\nu)} + \frac{1}{3} \d_{\a}{}^{(\mu} \d_{\b}{}^{\nu)} R + \frac{1}{6} R g_{\a\b} g^{\mu\nu} \ .
\eea
The tensor $M_{\a\b\mu\nu}$ is separately symmetric in the first and the second set of indices, and $M_{\a\b\mu\nu}=M_{\mu\nu\a\b}$.  Equation \re{eom2} has the correct limit around the Friedmann--Lema\^{\i}tre--Robertson--Walker metric, unlike putting $\d T_{\a\b}=0$, although the difference is only important for wavelengths close to or larger than the Hubble length (for zero spatial curvature). It can be compared to the equation of motion for the vector potential $A^\a$ of the electromagnetic field in the Lorenz gauge in curved spacetime discussed in \cite{Koksbang:2021alx}, $\Box A^{\a} - R^\a{}_\b A^\b = 0$. In the gravitational wave case, the Weyl tensor enters in addition to the Ricci tensor, and unlike for electromagnetic waves, direct coupling to the Riemann tensor is present in all gauges.

\para{The post-geometrical approximation.}

The geometrical optics approximation is based on the assumption that the field has the local plane wave form \cite{Isaacson:1968hbi}, \cite{Maggiore:2007ulw} (page 42)
\begin{align} \label{hexpansion}
	\bar h^{\a\b} = \sum_{n=0}^\infty \text{Re} ( \bar h^{\a\b}_n \epsilon^n e^{i S/\epsilon} ) \ ,
\end{align}
where the amplitude is expanded in a series of the constant parameter $\epsilon\ll1$, which represents the ratio of the wavelength to other relevant scales that are considered large. When the curvature radius is much larger than the wavelength, the leading terms give geometrical optics, and subleading terms give energy- and polarisation-dependent propagation, including the gravitational Hall effect for both light \cite{Oancea:2019pgm, Oancea:2020khc, Frolov:2020uhn} and gravitational waves \cite{Andersson:2020gsj, Oancea:2022szu, Oancea:2023hgu}. As in \cite{Koksbang:2021alx}, we instead consider a post-geometrical optics approximation where the spacetime curvature radius is not assumed to be large compared to the wavelength. (The curvature was also included in \cite{Cusin:2019rmt}, where the Riemann tensor was taken to be a perturbative correction and only the vacuum case was considered.) Inserting the expansion \re{hexpansion} into the equation of motion \re{eom2} gives
\bea \label{eomexp}
    \!\!\!\!\!\!\!\!\! 0 &=& \sum_{n=0}^\infty \text{Re} \left[ \epsilon^n \left( - \epsilon^{-2} k^2 \bar h^{\a\b}_n + i \epsilon^{-1} \nabla_\mu k^\mu \bar h^{\a\b}_n + 2 i \epsilon^{-1} k^\mu \nabla_\mu \bar h^{\a\b}_n + M^{\a\b}{}_{\mu\nu} \bar h^{\mu\nu}_n \right) e^{i S/\epsilon} \right] \ ,
\eea
where $k_\a\equiv\pat_\a S$ and $k^2\equiv k^\a k_\a$. We take the Riemann term $M^{\a\b}{}_{\mu\nu}$ to be of order $\epsilon^{-2}$. Expanding the Lorenz gauge condition \re{Lorenz} in the same way, we have
\bea \label{Lorenzexp}
    0 &=& \sum_{n=0}^\infty \text{Re} \left[ \epsilon^n \left( \epsilon^{-1} i k_\mu \bar h^{\mu\a}_n + \nabla_\mu \bar h^{\mu\a}_n \right) e^{i S/\epsilon} \right] \ .
\eea

Considering the equation of motion \re{eomexp} at orders $\epsilon^{-2}$ and $\epsilon^{-1}$ gives
\bea
  \label{go1} k^2 a^{\a\b} - M^{\a\b}{}_{\mu\nu} a^{\mu\nu} &=& 0 \\
  \label{go2} k^2 b^{\a\b} - M^{\a\b}{}_{\mu\nu} b^{\mu\nu} &=& a_{\a\b} \nabla_\mu k^\mu + 2 k^\mu \nabla_\mu a_{\a\b} \ ,
\eea
where we have denoted $\bar h^{\a\b}_0\equiv a^{\a\b}$, $\bar h^{\a\b}_1\equiv i b^{\a\b}$. Expanding the Lorenz gauge condition \re{Lorenzexp} to orders $\epsilon^{-1}$ and $\epsilon^0$ gives
\bea
  \label{go3} k_\mu a^{\mu\a} &=& 0 \\
  \label{go4} k_\mu b^{\mu\a} &=& \nabla_\mu a^{\mu\a} \ .
\eea
This result is the same as in geometrical optics and shows that the wave is transverse.

\para{The energy-momentum tensor.}

Consider a timelike unit vector $u^\a$. Without loss of generality, we decompose the energy-momentum tensor as
\bea \label{Tab}
  T_{\a\b} = ( \rho + p ) u_\a u_\b + p g_{\a\b} + 2 q_{(\a} u_{\b)} + \Pi_{\a\b} \ ,
\eea
where $\rho$ is the energy density, $p$ is the pressure, $q_\a$ is the energy flux, and $\Pi_{\a\b}$ is the anisotropic stress. Both $q_\a$ and $\Pi_{\a\b}$ are orthogonal to $u^\a$, and $\Pi_{\a\b}$ is traceless and symmetric. Inserting \re{Tab} into the background Einstein equation \re{Einstein} we get
\bea \label{Rab}
  \Mpl^2 R_{\a\b} &=& ( \rho + p  ) u_\a u_\b + \ha ( \rho - p ) g_{\a\b} + 2 q_{(\a} u_{\b)} + \Pi_{\a\b} \ .
\eea

Inserting the definition of $M_{\a\b\mu\nu}$ from \re{M} and the Ricci tensor from \re{Rab} into the leading order equation of motion \re{go1}, we have
\bea \label{eoma}
  0 &=& \left( \Mpl^2 k^2 + \frac{2}{3} \rho \right) a_{\a\b} - 2 \Mpl^2 C_{\a\mu\b\nu} a^{\mu\nu} + 2 ( \rho + p ) u_{(\a} a_{\b)\mu} u^\mu + 2 q_{(\a} a_{\b)\mu} u^\mu + 2 u_{(\a} a_{\b)\mu} q^\mu \el
  && + 2 \Pi_{(\a}{}^\mu a_{\b)\mu} - \frac{1}{6} a^\mu{}_\mu ( \rho - 3 p ) g_{\a\b} \ .
\eea
In the local plane wave approximation, the change in the direction of the wave is slow compared to the scale given by the wavenumber. As we are interested in a situation where the curvature changes rapidly, in order for the approximation to be valid, the direction given by $a_{\a\b}$ cannot be locally related to the energy-momentum tensor. We therefore must have $q_\a=0$, $\Pi_{\a\b}=0$, $a_{\a\mu} u^\mu=0$, $a^\mu{}_\mu=0$. From $a_{\a\mu} u^\mu=0$ it follows that $a_{\a\b} a^{\a\b}>0$; we write $a_{\a\b}=a f_{\a\b}$, where $a>0$ and $f_{\a\b} f^{\a\b}=1$. As $a_{\a\b}$ is orthogonal to both $u^\a$ and $k^\a$, it spans a two-dimensional space. From $a^\mu{}_\mu=0$ it follows that $C_{AMBN} a^{AB} = C_{ABCD} f^{AC} f^{BD} a_{MN}$, where capital letters label directions orthogonal to $u^\a$ and $k^\a$. The components of the Weyl tensor along the polarisation plane thus generate a mass term given by the Newman--Penrose scalars, which is in general non-zero everywhere, not just inside matter. However, the components along $u^\a$ and $k^\a$ make \re{eoma} inconsistent, unless they are so small that they can be neglected. For the Gaussian density profile we consider, this is the case, as we discuss in \sec{sec:disc} and appendix \ref{app:turn}. So \re{eoma} reduces to
\bea \label{mass}
  k^2 &=& - \frac{2 \rho}{3 \Mpl^2} \ .
\eea
Inside matter the effective graviton mass squared is negative (assuming $\rho>0$) so the tangent vector $k^\a$ is timelike, and gravitational waves slow down. The dispersion relation for a graviton with energy $E$ is
\bea
  E^2 = \vec k^2 + \frac{2 \rho}{3 \Mpl^2} \ ,
\eea
where $|\vec k|$ is the spatial wavenumber, and as usual the wavelength is $2\pi/|\vec k|$. (For a summary of similar dispersion relations for gravitational waves in matter found in various approximations, see \cite{Gayer:1979ff}; \cite{Ignatev:1974} is closest to the present work.) For $E^2< \frac{2 \rho}{3 \Mpl^2}$, the wave cannot propagate inside matter, and the local plane wave approximation is not valid. The situation is different from the case of massive gravitons considered in modified gravity: here the mass is non-zero only in the fraction of space where matter energy density is non-zero \cite{Colangeli:2025hrs}. Thus the gradient of the mass is non-zero, leading to non-geodesic motion.

Taking the covariant derivative of \re{mass} and using the fact that $\nabla_\b k_\a=\nabla_\a k_\b$ (since $k_\a=\pat_\a S$), we get
\bea \label{knk}
  k^\b \nabla_\b k_\mu &=& \ha \nabla_\mu k^2 = - \frac{1}{3 \Mpl^2} \pat_\mu \rho \ ,
\eea
showing that the curvature due to matter pushes gravitons off geodesics. The results \re{mass} and \re{knk} differ from the electromagnetic case considered in \cite{Koksbang:2021alx} only in that here the gravitationally induced mass squared is $\frac{2\rho}{3 \Mpl^2}$ instead of $\frac{\rho - p}{2 \Mpl^2}$. Therefore the effects on the distance are also similar, with only small changes in numerical factors. One might worry that large density gradients would cause the ray to turn strongly, invalidating the local plane wave approximation. We show in appendix \ref{app:turn} that this is not the case: as long as $\rho/(\Mpl^2 E^2)\ll1$, the change of the wave front direction is slow compared to the change of the phase.

Contracting \re{go2} with $a^{\a\b}$ and using \re{go1} and the property $M_{\a\b\mu\nu}=M_{\mu\nu\a\b}$, we obtain
\bea \label{gn}
    0 &=& \nabla_\mu ( a^2 k^\mu ) \ ,
\eea
so graviton number is conserved, as in geometrical optics. Using this decomposition in \re{go2} and applying \re{gn}, we find
\bea \label{pol}
  \! \! \! \! \! \! \! \! \! \! \! \! \! \! \! \! k^\mu \nabla_\mu f_{\a\b} &=& \frac{1}{2 a} ( k^2 b_{\a\b} - M_{\a\b\mu\nu} b^{\mu\nu} ) \el
  &=& - \frac{1}{a} C_{\a\mu\b\nu} b^{\mu\nu} +  \frac{1}{\Mpl^2 a} ( \rho + p ) u_{(\a} b_{\b)\mu} u^\mu - \frac{b^\mu{}_\mu}{12 \Mpl^2 a}( \rho - 3 p ) g_{\a\b} \ ,
\eea
where on the second line we have used \re{M}, \re{Rab}, and \re{mass}. The polarisation vector is not in general parallel transported along the light ray, but exhibits circular birefringence. It has been argued that the Weyl term generates extra polarisation modes even in vacuum when treated perturbatively as a higher-order term in $\epsilon$ \cite{Cusin:2019rmt, Dalang:2021qhu}. This is similar to how an electromagnetic field propagating through a medium can generate the scalar component $A^0$ even if it is initially set to zero. In neither case does the number of degrees of freedom grow: any new components are generated by the two propagating modes only. Understanding propagation beyond geometrical optics is important for observationally distinguishing between such possible novel effects in general relativity and modified gravity that can involve new physical degrees of freedom \cite{Menadeo:2024uoq, Menadeo:2025hgf}.

\subsection{Angular diameter distance} \label{sec:distance}

\para{Effect on the integrated area expansion rate.}

Repeating the calculation of \cite{Koksbang:2021alx} with \re{mass} and \re{knk} gives the post-geometrical correction to redshift and angular diameter distance. We sketch only the main points, and refer to \cite{Koksbang:2021alx} for the notation and details. As in the case of photons, the post-geometrical correction to the graviton redshift is negligible because the fraction of the volume occupied by particles is tiny. However, this is not necessarily true for the angular diameter distance, which involves derivatives of the density. As shown in \cite{Koksbang:2021alx}, to excellent accuracy we have $D_A\propto\exp\left( \ha\int\rmd\lambda \thetat \right)$ as in geometrical optics, where $\thetat$ is the area expansion rate of the ray bundle orthogonal to $k^\a$, $\lambda$ is the affine parameter along the ray, and
\bea \label{thetaint}
\int\rmd\lambda \thetat &=& \int\rmd\lambda \int\rmd\lambda' \left( - \frac{1}{2} \thetat^2 - 2 \sigmat^2 -R_{\a\b} k^{\a}k^{\b} + \frac{1}{2}\Box k^2 - P_{\a\b} P^{\a\b} \right) \ ,
\eea
where $\sigmat$ and $P_{\a\b}$ are defined via (2.31) of \cite{Koksbang:2021alx}. We have
\bea
  && \int\rmd\lambda P_{\a\b} P^{\a\b} \el
  &=& \int\rmd\lambda \sum_n \frac{1}{E^2 v^2} \left[ \frac{5}{12} ( \pat_{r_n} k^2)^2 + \frac{5}{3} \frac{E^2}{\rho} \pat_{r_n} k^2 \, \pat_{r_n} p - \frac{1}{3} k^2 \frac{4 E^2 - k^2}{\rho^2} ( \pat_{r_n} p )^2 \right] \ .
\eea
In the electromagnetic case, $k^2=-\frac{\rho-p}{2\Mpl^2}$ leads to (2.44) of \cite{Koksbang:2021alx}. For the gravitational wave result $k^2=-\frac{2\rho}{3\Mpl^2}$, we get instead
\bea \label{finalPP}
  \int\rmd\lambda P_{\a\b} P^{\a\b} &=& \int\rmd\lambda \sum_n \frac{1}{\Mpl^4 E^2 v^2} \left[ \frac{5}{27} ( \pat_{r_n} \rho)^2 - \frac{10}{9} \frac{\Mpl^2 E^2}{\rho} \pat_{r_n} \rho \, \pat_{r_n} p \right. \el
  && \left. + \frac{4}{27} \left( 1 + \frac{6 \Mpl^2 E^2}{\rho} \right) ( \pat_{r_n} p )^2 \right] \ ,
\eea

Adding up the terms in \re{thetaint} and assuming that the density distribution inside a single particle is a Gaussian with the width equal to the Compton wavelength, we obtain for gravitational waves (dropping small terms as in \cite{Koksbang:2021alx})
\bea \label{thetaint2}
  \int\rmd\lambda \thetat &=& \int\rmd\lambda \int\rmd\lambda' \left\{ - \frac{1}{2}\tilde{\theta}^2 - 2\sigmat^2 - E^2 \frac{\av{\rho}}{\Mpl^2} \left[ 1 + \frac{5}{4608\pi^{13/2}} \frac{m^6}{\Mpl^2 E^4} \right. \right. \el
  && \left. \left. - \frac{13}{36\sqrt{2}} \frac{\rho_{\text{max}}}{\Mpl^2 E^2} + \frac{5}{81\sqrt{3}} \left( \frac{\rho_{\text{max}}}{\Mpl^2 E^2} \right)^2 \right] \right\} \ ,
%  && \left. \left. - \frac{13}{108\sqrt{\pi}} \frac{\rho_0}{\Mpl^2 E^2} + \frac{10}{729\sqrt{3}\pi} \left( \frac{\rho_0}{\Mpl^2 E^2} \right)^2 \right] \right\} \ ,
\eea
%where $\rho_0=3m^4/(32\pi^4)$ is the mean density of a particle.
%
where $\rho_{\text{max}}=m^4/(2\pi)^{9/2}$ is the maximum density inside a particle. There are only small numerical differences in the coefficients compared to the photon result found in \cite{Koksbang:2021alx}. The maximum induced mass squared is $2\rho_{\text{max}}/(3\Mpl^2)$. If $E^2$ is smaller than this, the wave cannot propagate through the particles. For this limiting energy, the last two terms in \re{thetaint2} reach their maximum absolute value 0.4 and 0.08, respectively. So these correction terms are small unless the energy is below or close to the curvature scale, where the local plane wave approximation cannot be valid anymore.

\subsection{Discussion} \label{sec:disc}

\para{Effect on the distance.}

The relative correction to the angular diameter distance $D_A$ from \re{thetaint2} is $\frac{5}{4608\pi^{13/2}} \frac{m^6}{\Mpl^2 E^4}=6\times10^{-7}\frac{m^6}{\Mpl^2 E^4}$. (The equation for $D_A$ is (2.49) of \cite{Koksbang:2021alx}, with a different numerical coefficient for the curvature term.) This becomes of order unity for wave energy $E_D=0.03 m (m/\Mpl)^{1/2}=0.02 \text{ eV} (m/\text{GeV})^{3/2}$, or equivalently for particle mass $m=10 (E/\text{eV})^{2/3}$ GeV $=4(f/\text{Hz})^{2/3}$ eV, where $f$ is the frequency of the gravitational wave. The maximum curvature is of the order of the maximum graviton mass due to a particle, $E_{\text{C}}\equiv\sqrt{2\rho_{\text{max}}/(3\Mpl^2)}=0.01 m^2/\Mpl=5\times10^{-12}(m/\text{GeV})^2$ eV. Equivalently, the particle mass for which the curvature scale is equal to the graviton energy is $m=30(f/\text{Hz})^{1/2}$ MeV. In general, wave optics effects are important when the wavelength $2\pi/E$ (assuming $E\gg E_{\text{C}}$) is larger than the Schwarzschild radius $r_{\text{s}}=m/(4\pi\Mpl^2)$. The limiting energy is $E_{\text{WO}}=8\pi^2 \Mpl^2/m$. We denote $\varepsilon\equiv m/\Mpl$; as we consider particles such that classical gravity is a good approximation, $\varepsilon\ll1$. So the hierarchy of scales is $E_{\text{WO}}=80 m\varepsilon^{-2}\gg E_D=0.03 m \varepsilon^{1/2}\gg E_{\text{C}}=0.01 m \varepsilon$. The gravitational waves impacted by a single particle at a time are always deep in the wave optics regime, since the Schwarzschild radius goes down with decreasing mass. However, the small mass also means that the deflection angle is tiny. The graviton mass remains small, $\sim\varepsilon^{1/2} E$, even when the correction to the distance becomes significant, so the local plane wave approximation is not obviously wrong.

Gravitational wave energy is $E=4\times10^{-15} f$/Hz eV. For the local plane wave approximation to be valid, $E\gg E_{\text{C}}$ is necessary. The lowest LIGO--Virgo--KAGRA frequency $\sim10$ Hz corresponds to $E=4\times10^{-14}$ eV, so the approximation is reliable for $m\lesssim$ 100 MeV, and the distance correction is large for $m\gtrsim 10$ eV. For LISA, the smallest frequency is $\sim10^{-4}$ Hz, so the corresponding mass range is 0.1 MeV $\gtrsim m\gtrsim0.01$ eV. Pulsar timing arrays with the lowest frequency $\sim10^{-9}$ Hz correspond to the mass range 1 keV $\gtrsim m\gtrsim10^{-6}$ eV, but there is a measurable signal only if distance to the sources can be determined. In \cite{Koksbang:2021alx}, dark matter masses $10^4$ GeV $\gtrsim m\gtrsim100$ MeV were excluded by lack of observed CMB distortion.

In contrast to \cite{Koksbang:2021alx}, where only dark matter was relevant as light cannot propagate freely through charges, for gravitational waves also electrons and baryons have to be taken into account. Their correction to the distance is weighed by their contribution to the total matter density, which is $0.2$ for baryons and $10^{-5}$ for electrons. However, as the correction is proportional to $E^{-4}$, this has only a small effect on the limiting energy. Baryon masses are above the regime of validity for LIGO--Virgo--Kagra frequencies (and obviously for all smaller frequencies), so the post-geometrical optics approximation is clearly not valid, and no conclusions can be drawn. In contrast, electron mass is so small that the induced graviton mass remains negligible at the lowest frequency observed by LIGO--Virgo--KAGRA, so the local plane wave approximation is not obviously incorrect. The resulting correction to $D_A$ is overwhelmingly large. From (2.49) of \cite{Koksbang:2021alx} (replacing the numerical coefficient with the one derived here for gravitational waves), the equation of motion for small redshifts $z\ll1$ is $\frac{\rmd^2 D_A}{\rmd z^2}=-\b^2 D_A$, where $\b^2\equiv \frac{5}{3072\pi^{13/2}}\Omega_{e0}\frac{m_e^6}{\Mpl^2 E^4}=3\times10^{22}$, where $\Omega_{e0}$ is the electron density parameter, $m_e$ is electron mass, and $E=4\times10^{-14}$ eV corresponds to 10 Hz. This leads to caustic formation at distance $\pi \b^{-1} H_0^{-1} \sim0.1$ pc (where $H_0$ is the Hubble constant), well below any cosmological scales. The local plane wave approximation and hence the post-geometrical approximation is thus strongly violated.  (The distance is so small that the approximations leading to (2.49) of \cite{Koksbang:2021alx} that assume the cosmic average mass density are not valid, but this makes little difference to the conclusion.) Such caustics could have a strong effect on the distance-redshift relation on the LIGO--Virgo--KAGRA observations, but work beyond the post-geometrical optics approximation is needed to check their existence and impact.

\para{Dependence on density profile.}

The result depends on the shape of the density profile. However, changing the mass distribution while keeping fixed the scale over which it is distributed can only significantly increase the effect, not bring it down. The leading term in \re{finalPP} is proportional to $\int\rmd^3 r (\pat_r \rho)^2$, and we can obtain a lower limit for it as follows. If we approximate that the density goes to zero at radius $R$ (corresponding to volume $V$), we have $\frac{1}{V} \int\rmd^3 r (\pat_r \rho)^2 \geq \frac{1}{V^2} ( \int\rmd^3 r \pat_r \rho)^2$ from the non-negativity of variance. By partial integration and using $R\geq r$ inside the integral, we then obtain the limit $\int\rmd^3 r (\pat_r \rho)^2\geq 3 m^2/(\pi R^5)$. (A Gaussian of width $R$ goes under this limit by a factor of $16\sqrt{\pi}$.) There is no upper limit, as the integral diverges for a top-hat profile.

The result is strongly dependent on the scale on which the particles are localised. As noted in \cite{Koksbang:2021alx}, taking the de Broglie wavelength instead of the Compton wavelength as the width of the Gaussian has only a small effect. It changes the mass at which the effect on the distance is significant by the factor $v^{-5/6}$, where $v$ is particle velocity. For dark matter with $v\sim10^{-3}$ this increases the mass by $\sim10^2$, and for electrons with thermal velocities $v\sim10^{-2}$ (corresponding to temperature $\sim10^6$ K in the intergalactic medium) by $\sim10\ldots100$. From the quantum mechanical point of view, the localisation scale is determined by the interactions of the particle, and need not agree with the Compton nor de Broglie wavelength. Protons and neutrons, unlike elementary particles, have a lower limit on the width of their mass distribution set by the compositeness scale (which is roughly the same as the Compton wavelength), but the width can be much larger if the particle is not well localised. Charged particles are efficiently localised by interaction with the CMB, but it has been argued that the localisation scale of electrons is still macroscopic \cite{Joos:1984uk, Tegmark:1993yn}. This would reduce their curvature radius well below the range relevant for gravitational waves or electromagnetic waves. Dark matter particles should be less localised due to weaker interactions, but their case has been less studied \cite{Allali:2020ttz, Allali:2020shm, Allali:2021puy}. Determining the localisation scale requires further work.

\para{Weyl tensor.}

An important feature for gravitational waves which does not appear in the electromagnetic field case discussed in \cite{Koksbang:2021alx} is the presence of the Weyl term \re{eom2}, which pushes the waves away from null geodesics even in vacuum. (In curved spacetime, the connection coefficients inside the $\Box$ operator also push the wave away from null geodesics.) We have neglected the Weyl tensor. Its contribution to the effective mass averages out when integrating terms linear in the density, but not in the quadratic terms. For the Gaussian density profile, the Weyl tensor part is subdominant to the $(\pat_r\rho)^2$ term except in the region far from the centre, which gives only a small contribution to the total. When the correction to the distance starts to be large, the effective graviton mass remains negligible, as we have seen, so the part of the Weyl tensor that would change the direction of the wave (and violate the local plane wave approximation) is also small.

\section{Conclusions} \label{sec:conc}

\para{Post-geometrical optics and gravitational wave propagation.}

We have applied the post-geometrical optics approximation that was introduced for electromagnetic waves in \cite{Koksbang:2021alx} to gravitational waves. It consists of keeping the local plane wave approximation used in geometrical optics but retaining the curvature terms, which can be locally large due to the discrete nature of matter particles. The plane wave approximation only works if the gravitationally induced mass remains smaller than the energy of the wave. We have checked that in this case the change in wavefront direction also remains small, consistent with the approximation.

Within this regime of validity, if the particle mass is large enough, the curvature caused by it will push the waves off null geodesics and can significantly impact the angular diameter distance. Assuming that the mass distribution is a Gaussian with width equal to the Compton wavelength, in the electromagnetic case, the mass range where the approximation is valid and there would be a large effect on the CMB is $10^4$ GeV $\gtrsim m\gtrsim100$ MeV \cite{Koksbang:2021alx}. We have now found that for the frequency of the gravitational waves detected by the LIGO--Virgo--KAGRA collaboration, the corresponding mass range for which the approximation is valid and the effect on the distance is large is 100 MeV $\gtrsim m\gtrsim$ 10 eV.

The curvature can also lead to birefringence of gravitational waves, but its magnitude cannot be evaluated within the leading orders of the post-geometrical approximation, as the hierarchy of the equations does not close for the polarisation. Unlike in the electromagnetic case, the gravitational wave equation of motion includes not only the Ricci tensor but also the Weyl tensor, which affects the wave even in vacuum, but its effect is subdominant.

Gravitational waves, unlike electromagnetic radiation, are able to freely pass through charged matter. Baryons are outside the mass range relevant for LIGO--Virgo--KAGRA, but electrons are not, and they would strongly focus gravitational waves on distances much smaller than cosmological scales. This leads to caustics on very small scales, invalidating the post-geometrical optics approximation. Therefore, the regime of validity of the post-geometrical approximation seems limited, as it stops being applicable when the curvature effects become large. This also applies to the electromagnetic case considered in \cite{Koksbang:2021alx}. An improved treatment is needed to account for the curvature spikes that particle matter can produce in the equation of motion of gravitational waves and electromagnetic radiation, and whether they lead to caustics. More work on the decoherence and localisation of particles is also needed to better understand the localisation scale of the mass distribution of individual particles.

\acknowledgments

SR thanks Asta Heinesen and Mark Hindmarsh for helpful discussions, and Laboratoire de Physique de l’Ecole Normale Sup\'erieure for hospitality.

\appendix

\section{Change in wavefront direction} \label{app:turn}

In this appendix we show that the turning of the ray caused by the curvature of a single particle is small as long as the induced gravitational mass \re{mass} is small, showing that this aspect of the local plane wave approximation is consistent. The analysis is the same in the present gravitational wave case and the electromagnetic case studied in \cite{Koksbang:2021alx}. We write $k^2=-\frac{2\kappa\rho}{\Mpl^2}$ to treat both at the same time, with $\kappa=1/3$ for gravitons and $\kappa=1/4$ for photons (the small pressure contribution in the photon case makes no difference and can be neglected). The modified geodesic equation \re{knk} reads
\bea
  k^\b \nabla_\b k_\mu &=& - \frac{\kappa}{\Mpl^2} \pat_\mu \rho \ .
\eea
We drop the effect of the gravitational potential generated by the particle, as it is negligible because of the small mass. We parametrise the ray orthogonal to the wavefront as $x^\a(\lambda)=(t,x^i)=(t,x,y,z)$, with $k^\a=\frac{\rmd x^\a}{\rmd\lambda}$. We then have
\bea \label{acc}
  \frac{\rmd^2 x_\a}{\rmd\lambda^2} = - \frac{\kappa}{\Mpl^2} \pat_\a \rho(x,y,z) \ .
\eea
As the density does not depend on time, the $\a=0$ component simply gives $t=E\lambda$, where $E$ is the constant wave energy, and we have without loss of generality set $t(0)=0$. (We work in the regime $\rho/(\Mpl^2 E^2)\ll1$, where the plane wave approximation can be valid, so the wave energy is constant to leading order.) Following \cite{Koksbang:2021alx}, we consider a Gaussian density profile of width $R$,
\bea \label{rho}
  \rho(x^i) &=& \frac{m}{(2\pi)^{3/2} R^3} e^{-\frac{r^2}{2 R^2}} \ ,
\eea
where $r\equiv\d_{ij} x^i x^j$. We solve \re{acc} perturbatively, $x^i=\bar x^i+\d x^i$. By symmetry the motion is restricted to a plane, so the leading order solution can without loss of generality be written as
\bea
  \bar x^i = ( E \lambda, b, 0 ) \ ,
\eea
where $b$ is the impact parameter. The equation for the perturbation is then
\bea
  \frac{\rmd^2 \d x^i}{\rmd\lambda^2} = \bar x^i \frac{\kappa m}{(2\pi)^{3/2} R^5 \Mpl^2} e^{-\frac{E^2\lambda^2 + b^2}{2 R^2}} \ ,
\eea
with the solution (given the initial condition $\d x^i(-\infty)=0$)
\bea
  \d x &=& - R \frac{ \kappa \rho(0,b,0)}{\Mpl^2 E^2} \sqrt{\frac{\pi}{2}} \left[ 1+ \erf\left( \frac{t}{\sqrt{2} R} \right) \right] \el
  \d y &=&  b \frac{ \kappa \rho(0,b,0)}{\Mpl^2 E^2} \left\{ e^{-\frac{t^2}{2 R^2}} + \frac{\sqrt{\pi}t}{\sqrt{2} R} \left[ 1+ \erf\left( \frac{t}{\sqrt{2} R} \right) \right] \right\} \ ,
\eea
where $\erf(x)$ is the error function. The non-zero components of $k^i=\bar k^i +\d k^i$ are thus
\bea
  k^x &=& E \left( 1 - \frac{ \kappa \rho(0,b,0)}{\Mpl^2 E^2} e^{-\frac{t^2}{2 R^2}} \right) \el
  k^y &=& E \sqrt{\frac{\pi}{2}} \frac{b}{R} \frac{ \kappa \rho(0,b,0)}{\Mpl^2 E^2} \left[ 1+ \erf\left( \frac{t}{\sqrt{2} R}\right) \right] \ .
\eea
The deflection angle $\theta=\arcsin(k^y/k^x)$ is then (working to first order in $\frac{\rho(0,b,0)}{\Mpl^2 E^2}$) smaller than $\kappa\sqrt{2\pi}e^{-1/2}\frac{\rho_{\text{max}}}{\Mpl^2 E^2}$. So even though the gradient of the density enhances the turning rate (for $R=\lc$) by the large factor $m$, integration smooths this, and the turning of the ray is small for $\frac{\rho_{\text{max}}}{\Mpl^2 E^2}\ll1$. This means that the direction of the wavefront does not significantly change when passing through particles, which is a necessary requirement for the local plane wave approximation to be valid. As the Weyl term is subdominant to the density term, this also shows that its effect on wavefront direction in \re{eoma} can be neglected.

\bibliographystyle{JHEP}
\bibliography{disc}

\providecommand{\href}[2]{#2}\begingroup\raggedright\begin{thebibliography}{10}

\bibitem{Isaacson:1968hbi}
R.~A. Isaacson, \emph{{Gravitational Radiation in the Limit of High Frequency.
  I. The Linear Approximation and Geometrical Optics}},
  \href{http://dx.doi.org/10.1103/PhysRev.166.1263}{\emph{Phys. Rev.}
  {\bfseries 166} (1968) 1263--1271}.

\bibitem{Maggiore:2007ulw}
M.~Maggiore, \emph{{Gravitational Waves. Vol. 1: Theory and Experiments}}.
\newblock Oxford University Press, 2007,
  \href{http://dx.doi.org/10.1093/acprof:oso/9780198570745.001.0001}{10.1093/acprof:oso/9780198570745.001.0001}.

\bibitem{Ohanian:1973}
H.~C. Ohanian, \emph{{Focusing of Gravitational Radiation}},
  \href{http://dx.doi.org/10.1103/PhysRevD.8.2734}{\emph{Phys. Rev. D}
  {\bfseries 8} (1973) 2734--2735}.

\bibitem{Ohanian:1974}
H.~C. Ohanian, \emph{{On the Focusing of Gravitational Radiation}},
  \href{http://dx.doi.org/10.1007/BF01810927}{\emph{Int. J. of Theor. Phys.}
  {\bfseries 9} (1974) 425--437}.

\bibitem{Braga:2024pik}
G.~Braga, A.~Garoffolo, A.~Ricciardone, N.~Bartolo and S.~Matarrese,
  \emph{{Proper time path integrals for gravitational waves: an~improved wave
  optics framework}},
  \href{http://dx.doi.org/10.1088/1475-7516/2024/11/031}{\emph{JCAP} {\bfseries
  11} (2024) 031}, [\href{https://arxiv.org/abs/2405.20208}{{\ttfamily
  2405.20208}}].

\bibitem{Lanczos:1925}
K.~Lanczos, \emph{{Zum Problem der unendlich schwachen Felder in der
  Einsteinschen Gravitationstheorie}}, {\emph{Zeitschrift f{\"u}r Physik}
  {\bfseries 31} (1925) 112--132}.

\bibitem{Peters:1966}
P.~Peters, \emph{{Perturbations in the Schwarzschild Metric}},
  \href{http://dx.doi.org/10.1103/PhysRev.146.938}{\emph{Phys. Rev.} {\bfseries
  146} (1966) 938--946}.

\bibitem{Koksbang:2021alx}
S.~M. Koksbang and S.~R{\"a}s{\"a}nen, \emph{{The effect of dark matter
  discreteness on light propagation}},
  \href{http://dx.doi.org/10.1088/1475-7516/2022/04/030}{\emph{JCAP} {\bfseries
  04} (2022) 030}, [\href{https://arxiv.org/abs/2108.06163}{{\ttfamily
  2108.06163}}].

\bibitem{Gayer:1979ff}
S.~Gayer and C.~F. Kennel, \emph{{Possibility of Landau damping of
  gravitational waves}},
  \href{http://dx.doi.org/10.1103/PhysRevD.19.1070}{\emph{Phys. Rev. D}
  {\bfseries 19} (1979) 1070--1083}.

\bibitem{Ehlers:1987nm}
J.~Ehlers, A.~R. Prasanna and R.~A. Breuer, \emph{{Propagation of Gravitational
  Waves Through Pressureless Matter}},
  \href{http://dx.doi.org/10.1088/0264-9381/4/2/009}{\emph{Class. Quant. Grav.}
  {\bfseries 4} (1987) 253--264}.

\bibitem{Ehlers:1996}
J.~Ehlers and A.~R. Prasanna, \emph{{A WKB formalism for multicomponent fields
  and its application to gravitational and sound waves in perfect fluids}},
  \href{http://dx.doi.org/10.1088/0264-9381/13/8/016}{\emph{Class. Quant.
  Grav.} {\bfseries 13} (1996) 2231--2240}.

\bibitem{Oancea:2019pgm}
M.~A. Oancea, C.~F. Paganini, J.~Joudioux and L.~Andersson, \emph{{An overview
  of the gravitational spin Hall effect}},
  \href{https://arxiv.org/abs/1904.09963}{{\ttfamily 1904.09963}}.

\bibitem{Oancea:2020khc}
M.~A. Oancea, J.~Joudioux, I.~Y. Dodin, D.~E. Ruiz, C.~F. Paganini and
  L.~Andersson, \emph{{Gravitational spin Hall effect of light}},
  \href{http://dx.doi.org/10.1103/PhysRevD.102.024075}{\emph{Phys. Rev. D}
  {\bfseries 102} (2020) 024075},
  [\href{https://arxiv.org/abs/2003.04553}{{\ttfamily 2003.04553}}].

\bibitem{Frolov:2020uhn}
V.~P. Frolov, \emph{{Maxwell equations in a curved spacetime: Spin optics
  approximation}},
  \href{http://dx.doi.org/10.1103/PhysRevD.102.084013}{\emph{Phys. Rev. D}
  {\bfseries 102} (2020) 084013},
  [\href{https://arxiv.org/abs/2007.03743}{{\ttfamily 2007.03743}}].

\bibitem{Andersson:2020gsj}
L.~Andersson, J.~Joudioux, M.~A. Oancea and A.~Raj, \emph{{Propagation of
  polarized gravitational waves}},
  \href{http://dx.doi.org/10.1103/PhysRevD.103.044053}{\emph{Phys. Rev. D}
  {\bfseries 103} (2021) 044053},
  [\href{https://arxiv.org/abs/2012.08363}{{\ttfamily 2012.08363}}].

\bibitem{Oancea:2022szu}
M.~A. Oancea, R.~Stiskalek and M.~Zumalac{\'a}rregui, \emph{{Frequency- and
  polarization-dependent lensing of gravitational waves in strong gravitational
  fields}}, \href{http://dx.doi.org/10.1103/PhysRevD.109.124045}{\emph{Phys.
  Rev. D} {\bfseries 109} (2024) 124045},
  [\href{https://arxiv.org/abs/2209.06459}{{\ttfamily 2209.06459}}].

\bibitem{Oancea:2023hgu}
M.~A. Oancea, R.~Stiskalek and M.~Zumalac{\'a}rregui, \emph{{Probing general
  relativistic spin{\textendash}orbit coupling with gravitational waves from
  hierarchical triple systems}},
  \href{http://dx.doi.org/10.1093/mnrasl/slae084}{\emph{Mon. Not. Roy. Astron.
  Soc.} {\bfseries 535} (2024) L1--L6},
  [\href{https://arxiv.org/abs/2307.01903}{{\ttfamily 2307.01903}}].

\bibitem{Cusin:2019rmt}
G.~Cusin and M.~Lagos, \emph{{Gravitational wave propagation beyond geometric
  optics}}, \href{http://dx.doi.org/10.1103/PhysRevD.101.044041}{\emph{Phys.
  Rev. D} {\bfseries 101} (2020) 044041},
  [\href{https://arxiv.org/abs/1910.13326}{{\ttfamily 1910.13326}}].

\bibitem{Ignatev:1974}
Y.~Ignat'ev, \emph{{Dispersion of gravitational waves in a relativistic gas}},
  \href{http://dx.doi.org/10.1007/BF00892891}{\emph{Soviet Physics Journal}
  {\bfseries 17} (1974) 1749--1753}.

\bibitem{Colangeli:2025hrs}
E.~Colangeli, C.~Dalang and T.~Baker, \emph{{Multi-messenger lensing time delay
  as a probe of the graviton mass}},
  \href{https://arxiv.org/abs/2509.03196}{{\ttfamily 2509.03196}}.

\bibitem{Dalang:2021qhu}
C.~Dalang, G.~Cusin and M.~Lagos, \emph{{Polarization distortions of lensed
  gravitational waves}},
  \href{http://dx.doi.org/10.1103/PhysRevD.105.024005}{\emph{Phys. Rev. D}
  {\bfseries 105} (2022) 024005},
  [\href{https://arxiv.org/abs/2104.10119}{{\ttfamily 2104.10119}}].

\bibitem{Menadeo:2024uoq}
N.~Menadeo and M.~Zumalac{\'a}rregui, \emph{{Gravitational wave propagation
  beyond general relativity: Geometric optic expansion and lens-induced
  dispersion}},
  \href{http://dx.doi.org/10.1103/PhysRevD.111.104022}{\emph{Phys. Rev. D}
  {\bfseries 111} (2025) 104022},
  [\href{https://arxiv.org/abs/2411.07164}{{\ttfamily 2411.07164}}].

\bibitem{Menadeo:2025hgf}
N.~Menadeo, S.~Giardino and M.~Zumalac{\'a}rregui, \emph{{Gravitational-wave
  dispersion over inhomogeneous space-times: General relativity, screened
  theories of gravity and non-minimal dark energy}},
  \href{https://arxiv.org/abs/2511.08023}{{\ttfamily 2511.08023}}.

\bibitem{Joos:1984uk}
E.~Joos and H.~D. Zeh, \emph{{The Emergence of classical properties through
  interaction with the environment}},
  \href{http://dx.doi.org/10.1007/BF01725541}{\emph{Z. Phys. B} {\bfseries 59}
  (1985) 223--243}.

\bibitem{Tegmark:1993yn}
M.~Tegmark, \emph{{Apparent wave function collapse caused by scattering}},
  \href{http://dx.doi.org/10.1007/BF00662807}{\emph{Found. Phys. Lett.}
  {\bfseries 6} (1993) 571},
  [\href{https://arxiv.org/abs/gr-qc/9310032}{{\ttfamily gr-qc/9310032}}].

\bibitem{Allali:2020ttz}
I.~Allali and M.~P. Hertzberg, \emph{{Gravitational Decoherence of Dark
  Matter}}, \href{http://dx.doi.org/10.1088/1475-7516/2020/07/056}{\emph{JCAP}
  {\bfseries 07} (2020) 056},
  [\href{https://arxiv.org/abs/2005.12287}{{\ttfamily 2005.12287}}].

\bibitem{Allali:2020shm}
I.~J. Allali and M.~P. Hertzberg, \emph{{Decoherence from General Relativity}},
  \href{http://dx.doi.org/10.1103/PhysRevD.103.104053}{\emph{Phys. Rev. D}
  {\bfseries 103} (2021) 104053},
  [\href{https://arxiv.org/abs/2012.12903}{{\ttfamily 2012.12903}}].

\bibitem{Allali:2021puy}
I.~J. Allali and M.~P. Hertzberg, \emph{{General Relativistic Decoherence with
  Applications to Dark Matter Detection}},
  \href{http://dx.doi.org/10.1103/PhysRevLett.127.031301}{\emph{Phys. Rev.
  Lett.} {\bfseries 127} (2021) 031301},
  [\href{https://arxiv.org/abs/2103.15892}{{\ttfamily 2103.15892}}].

\end{thebibliography}\endgroup

\end{document}